\documentclass[twocolumn,preprintnumbers,amsmath,amssymb,nofootinbib]{revtex4}

\usepackage{graphicx}% Include figure files
\usepackage{epsfig}
\usepackage{color}
\usepackage{amsmath}

\newcommand{\beq}{\begin{equation}}
\newcommand{\eeq}{\end{equation}}
\newcommand{\bea}{\begin{eqnarray}}
\newcommand{\eea}{\end{eqnarray}}
\newcommand{\bwd}{\begin{widetext}}
\newcommand{\ewd}{\end{widetext}}

\begin{document}

\title{High Order Numerical Integrators for Relativistic Charged Particle Tracking}

%\author{Ji Qiang\thanks{jqiang@lbl.gov}, Lawrence Berkeley National Laboratry, Berkeley, USA }

\author{Ji Qiang}
\email{jqiang@lbl.gov}
\affiliation{Lawrence Berkeley National Laboratory, Berkeley, CA 94720, USA}

\begin{abstract}
In this paper, we extend several time reversible numerical integrators to
solve the Lorentz force equations from
second order accuracy to
higher order accuracy for relativistic charged particle tracking in 
electromagnetic fields.
A fourth order algorithm is given explicitly and tested with numerical 
examples. Such high order numerical integrators can significantly save the
computational cost by using a larger step size in comparison to the second
order integrators.

%can be useful for long-term tracking study of the space-charge effects in high intensity accelerators.
\end{abstract}

\maketitle

\section{Introduction}

Numerical tracking charged particle in electric and magnetic fields has 
many applications in beam physics and plasma physics.
It normally involves solving the Lorentz force equations numerically 
with external electromagnetic fields.
In previous studies, a second order, time reversible numerical 
algorithm, known as Boris integrator~\cite{boris} has been widely used in plasma
and beam physics simulations.
However, for numerical simulation of relativistic charged particle
motion in electromagnetic fields,
this integrator can produce large error~\cite{vay,qiangx,qiangy}. 
A new time reversible second order integrator was proposed in
reference~\cite{vay} that avoids this problem.
Recently, another time reversible second order integrator was proposed
by Higuera and Cary~\cite{hc}.
Besides working well for particle tracking in electromagnetic fields
with large relativistic factor, this
algorithm also preserves phase space volume.

The above three time reversible numerical integrators are second order
accuracy of integration step size.
In some numerical simulations, a higher order numerical integrator can
be more effective in attaining the desired numerical accuracy.
So far, the extension of these second order integrators to higher
order accuracies have not been reported in literature.
Meanwhile, in the area of symplectic numerical integrator study of
Hamiltonian systems, high order numerical integrators have been 
reported by using a split-operator method~\cite{forest1,yoshida}.
In this paper, after reformating the original Lorentz force equations,
we observed that higher order numerical integrators can be obtained
from these symmetric, time reversible second order integrators for
relativistic charge particle tracking.

The organization of this paper is as follows: after the introduction, we present the
high order numerical integrator in section II; 
We present
numerical tests of the fourth order integrator in Section III and
draw conclusions in Section IV.

\section{High Order Numerical Integrators}

The Lorentz equations of motion for a charged particle subject to electric and
magnetic fields can be written as:
\begin{eqnarray}
	\frac{d {\bf r}}{d t} & = & \frac{{\bf p}}{\gamma} \\
	\frac{d {\bf p}}{d t} & = & q(\frac{{\bf E}}{m c} + \frac{1}{\gamma}
	        {\bf p}\times {\bf B})
\end{eqnarray}
where ${\bf r} = (x,y,z)$ denotes the particle spatial coordinates, ${\bf p}=
(p_{x}/mc,p_{y}/mc,p_{z}/mc)$ the particle normalized mechanic momentum,
$m$ the particle rest mass, $q$ the particle charge, 
$c$ the speed of light in vacuum,
$\gamma$ the relativistic factor defined by $\sqrt{1+{\bf p}\cdot {\bf p}}$,
$t$ the time, ${\bf E}(x,y,z,t)$ the electric field, and ${\bf B}(x,y,z,t)$ the magnetic field.
%Here, both the electric field ($\bf E(x,y,z,t)$) 
%and the magnetic field are a function of
%spatial coordinates and time.
Instead of using the time $t$ as an explicit independent variable,
we write the above equations using $s$ as independent variable:
\begin{eqnarray}
	\frac{d t}{d s} & = & 1 \\
	\frac{d {\bf r}}{d s} & = & \frac{{\bf p}}{\gamma} \\
	\frac{d {\bf p}}{d s} & = & q(\frac{{\bf E}}{m c} + \frac{1}{m\gamma}
	        {\bf p}\times {\bf B})
\end{eqnarray}

Letting $\zeta(t,{\bf r},{\bf p}:s)$ denote a vector of coordinates,
the above equations of motion can be rewritten as:
\begin{eqnarray}
	\frac{d \zeta}{d s} & = & A \zeta 
\end{eqnarray}
where the matrix $A$ is a given as:
\begin{eqnarray}
	A & = & \begin{pmatrix}
	1/t & 0 & 0 \\
	0 & 0 & 1/\gamma \\
	0 & q {\bf E}/(mc{\bf r}) & q {\bf 1}\times {\bf B}/(m \gamma) 
	        \end{pmatrix}
\end{eqnarray}
A formal solution for above equation
after a single step $\tau$ can be written as:
\begin{eqnarray}
	\zeta (\tau) & = & \exp(A\tau) \zeta(0)
\end{eqnarray}
The matrix $A$ can be written as a sum of two terms $A = B + C$, where
\begin{eqnarray}
	B & = & \begin{pmatrix}
	1/t & 0 & 0 \\
	0 & 0 & 1/\gamma \\
	0 & 0 & 0 
	        \end{pmatrix}
\end{eqnarray}
and 
\begin{eqnarray}
	C & = & \begin{pmatrix}
	0 & 0 & 0 \\
	0 & 0 & 0 \\
	0 & q {\bf E}/(mc{\bf r}) & q {\bf 1}\times {\bf B}/(m \gamma) 
	        \end{pmatrix}
\end{eqnarray}
Using the Baker-Campbell-Hausdorff theorem~\cite{baker,campbell,haus},
a second order approximation for above single step 
solution can be obtained as:
\begin{eqnarray}
	\zeta (\tau) & = & \exp(\tau(B+C)) \zeta(0) \nonumber \\
				   & = & \exp(\frac{1}{2}\tau B)\exp(\tau C) \exp(\frac{1}{2}\tau B) \zeta(0) + O(\tau^3)
\end{eqnarray}
Letting $\exp(\frac{1}{2}\tau B)$ define a transfer map ${\mathcal M}_1$ and
$\exp(\tau C)$ a transfer map ${\mathcal M}_2$, 
for a single step, the above splitting results in a second order numerical integrator
for the original equation as:
\begin{eqnarray}
	\zeta (\tau) & = & {\mathcal M}(\tau) \zeta(0) \nonumber \\
    & = & {\mathcal M}_1(\tau/2) {\mathcal M}_2(\tau) {\mathcal M}_1(\tau/2) \zeta(0)
	+ O(\tau^3)
	\label{map}
\end{eqnarray}
From definitions of the matrices $B$ and $C$, it is seen that
the transfer map ${\mathcal M}_1$ corresponds to the solutions of 
Eqs.~3 and 4 for half step, and transfer map ${\mathcal M}_2$ corresponds to the 
solution of Eq.~5 for one step. 
%This type of algorithm is also known
%as the leap-frog algorithm for numerical integration~\cite{magenta}. 
The
solutions of transfer map ${\mathcal M}_1(\tau/2)$ is straightforward and can be 
written as:
\begin{eqnarray}
	t(\tau/2) & = & t(0) + \frac{\tau}{2} \\
	{\bf r}(\tau/2) & = & {\bf r}(0) + \frac{\tau\bf p}{2\gamma}
\end{eqnarray}
The solution for ${\mathcal M}_2(\tau)$ can have different forms depending
on different ways of approximation. In the Boris algorithm, 
${\mathcal M}_2(\tau)$ is given as:
\begin{eqnarray}
	{\bf p}_{-} & = & {\bf p}(0) + \frac{q {\bf E} \tau}{2mc} \\
	\gamma_{-} & = & \sqrt{1+{\bf p}_-\cdot {\bf p}_-} \\
	{\bf p}_+-{\bf p}_- & = & ({\bf p}_{+}+{\bf p}_{-}) \times 
\frac{q {\bf B}\tau}{2m\gamma_-}  \\
{\bf p}(\tau) & = & {\bf p}_+ + \frac{q {\bf E} \tau}{2mc} 
\end{eqnarray}
where ${\bf p}_+$ can be solved analytically from the linear equation
Eq.~17.
The Boris algorithm is time reversible and
has been widely used in numerical plasma and beam physics simulations.
However, it was found that the Boris algorithm could
have large numerical error for charged particle tracking when the
particle relativistic factor is large{vay,qiangx,qiangy}. The source of this
numerical error might result from the momentum update in separate steps from
the electric field and from the magnetic field.
This becomes especially
a serious problem to simulate a relativistic charged particle 
beam including space-charge effects, 
where the electric field and the magnetic field cancel each other
significantly in the laboratory frame and results in $1/\gamma^2$ decrease
of the transverse space-charge effects.
The new time reversible solution that avoids this problem
in ${\mathcal M}_2(\tau)$ was
proposed in reference~\cite{vay} as:
\begin{eqnarray}
	\gamma_0 & = & \sqrt{1+{\bf p}\cdot {\bf p}} \\
	{\bf p}_{-} & = & {\bf p}(0) + \frac{q \tau}{2mc}({\bf E} 
	+ c{\bf p}/\gamma_0 \times {\bf B}) \\
	{\bf p}_{+} & = & {\bf p}_- + \frac{q {\bf E} \tau}{2mc} \\
	\gamma_1 & = & \sqrt{1+{\bf p}_+\cdot {\bf p}_+} \\
	{\bf t} & = & \frac{q{\bf B}\tau}{2m} \\
        \lambda & = & {\bf p}_{+} \cdot {\bf t} \\
	\sigma & = & \gamma_1^2-{\bf t}\cdot{\bf t} \\
        \gamma_2 & = & \sqrt{\frac{\sigma+
\sqrt{\sigma^2+4({\bf t}\cdot{\bf t} + \lambda^2)}}{2}}  \\
{\bf t}^* & = & {\bf t}/\gamma_2  \\
s & = & 1/(1+{\bf t}^* \cdot {\bf t}^*)   \\
{\bf p}(\tau) & = & s[{\bf p}_+ + ( {\bf p}_+ \cdot{\bf t}^*){\bf t}^* +
{\bf p}_+ \times {\bf t}^*]
\end{eqnarray}
This algorithm does not have the problem of the Boris algorithm and
works well for relativistic particle tracking.
Recently, another time reversible and structure-preserving algorithm for
${\mathcal M}_2(\tau)$ was
proposed in reference~\cite{hc}. This algorithm is similar to 
the Boris algorithm by replacing the $\gamma_-$ in Eq.~17 with
the following $\gamma_{new}$:
\begin{eqnarray}
	{\bf p}_{-} & = & {\bf p}(0) + \frac{q {\bf E} \tau}{2mc} \\
	\gamma_{-} & = & \sqrt{1+{\bf p}_-\cdot {\bf p}_-} \\
	{\bf t} & = & \frac{q{\bf B}\tau}{2m} \\
	\gamma_{new} & = & \sqrt{\frac{\gamma_-^2-{\bf t} \cdot {\bf t}+
	 \sqrt{(\gamma_-^2-{\bf t} \cdot {\bf t})^2+4({\bf t} \cdot {\bf t}+|{\bf p}_- \cdot {\bf t}|^2}) }{2}}
\end{eqnarray}
This algorithm also works well for particle tracking with
large relativistic factor and
preserves phase space volume.

So far, all these three time reversible algorithms have a second order
accuracy of integration step size. The second order one step map can be rewritten as:
\begin{eqnarray}
	{\mathcal M}_{2nd}(\tau) & = & {\mathcal M}_1(\tau/2) {\mathcal M}_2(\tau) {\mathcal M}_1(\tau/2)
\end{eqnarray}
Since this one-step map is symmetric and time reversible, one can 
follow the exact steps of the reference~\cite{yoshida} and construct
the same fourth order accuracy numerical integrator as:
\begin{eqnarray}
	{\mathcal M}_{4th}(\tau) & = & {\mathcal M}_1(\frac{s}{2}) {\mathcal M}_2(s) {\mathcal M}_1(\frac{\alpha s}{2}) {\mathcal M}_2((\alpha-1)s)  \nonumber \\
					       & &	{\mathcal M}_1(\frac{\alpha s}{2}) {\mathcal M}_2(s) {\mathcal M}_1(\frac{s}{2})
	\label{map4}
\end{eqnarray}
where $\alpha = 1-2^{1/3}$, and $s=\tau/(1+\alpha)$.
An arbitrary even order accuracy integrator can also be obtained
following that reference.
%following Yoshida's approach~\cite{yoshida}.
Assume that ${\mathcal M}_{2n}$ denotes a transfer map with an accuracy of order $2n$, the transfer map
${\mathcal M}_{2n+2}$ with $(2n+2)$th order of accuracy can be obtained from the recursion equation~\cite{yoshida}:
\begin{eqnarray}
	{\mathcal M}_{2n+2}(\tau) & = & {\mathcal M}_{2n}(z_0\tau) {\mathcal M}_{2n}(z_1\tau) {\mathcal M}_{2n}(z_0 \tau) 
	\label{map2n}
\end{eqnarray}
where $z_0 = 1/(2-2^{1/(2n+1)})$ and $z_1 = -2^{1/(2n+1)}/(2-2^{1/(2n+1)})$.

\section{Numerical Tests}

\begin{figure}[!htb]
%\begin{figure}[htb]
%   \vspace*{-.5\baselineskip}
   \centering
   \includegraphics*[angle=270,width=200pt]{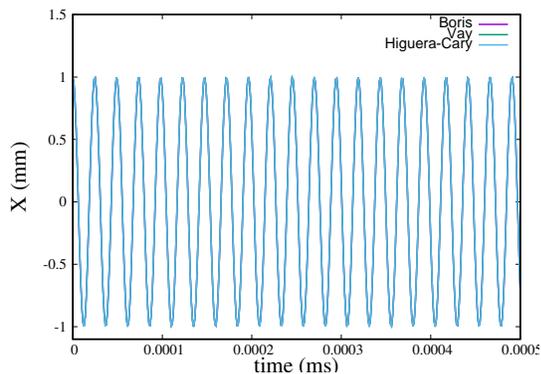}
   \caption{Particle trajectory evolution as a function of time from the
   $4^{th}$ order extension of the Boris integrator (magenta), the Vay integrator (green),
   and the Higuera-Cary integrator (blue) for an electron
   with $2$ MeV kinetic energy.}
   \label{fig1}
%   \vspace*{-\baselineskip}
\end{figure}

We tested the above $4^{th}$ order extension of the Boris
algorithm, the Vay algorithm, and the Higuera-Cary algorithm using two numerical
examples. In the first example, we considered an electron moving inside
static electric and magnetic fields. These fields are
given as:
\begin{eqnarray}
	E_x  & = & E_0 x \gamma_0 \\
	E_y  & = & E_0 y \gamma_0 \\
	E_z  & = & 0 \\
	B_x  & = & E_0 y (-\gamma_0 \beta_0/c)\\
	B_y  & = & E_0 x \gamma_0 \beta_0/c \\
	B_z  & = & 0 
\end{eqnarray}
where $\gamma_0$ is the relativistic factor of the moving beam,
$\beta_0 = \sqrt{1-(1/\gamma_0)^2}$, and the constant $E_0 = 9 \times 10^6$.
The above external fields correspond to 
the space-space fields generated by a moving
infinitely long transverse uniform cylindrical positron beam. 
First, we assume that both the initial electron kinetic energy 
and the moving positron beam kinetic energy
are $2$ MeV.
Figure~\ref{fig1} shows the electron trajectory evolution as a function of time from the
$4^{th}$ order extension of the Boris integrator (magenta), the Vay integrator (green),
and the Higuera-Cary integrator (blue) with a step size of $0.25$ ns (around $0.01$ oscillation period).
It is seen that three numerical integrators agree with each other very 
well in this case.
\begin{figure}[!htb]
%\begin{figure}[htb]
%   \vspace*{-.5\baselineskip}
   \centering
   \includegraphics*[angle=270,width=200pt]{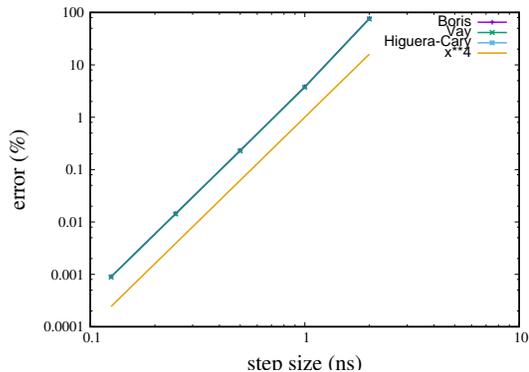}
   \caption{Relative numerical errors at the end of above integration
	   as a function of step size from the
   $4^{th}$ order extension of the Boris integrator (magenta), the Vay integrator (blue),
   and the Higuera-Cary integrator (green) for an electron
   with $2$ MeV kinetic energy. A power $4$ polynomial is also plotted here (orange).}
   \label{fig2}
%   \vspace*{-\baselineskip}
\end{figure}
Figure~\ref{fig2} shows the relative numerical errors at the end of integration
	   as a function of step size from the
   $4^{th}$ order extension of the Boris integrator (magenta), the Vay integrator (green),
   and the Higuera-Cary integrator (blue) together with a power $4$ polynomial.
It is seen that all three numerical integrators have nearly the same relative
errors and converge as $4^{th}$ power with respect to the step size.
Next, we assumed that both the initial electron and the moving positron beam
have a kinetic energy of $50$ MeV.
Figure~\ref{fig3} shows the electron trajectory evolution as a function of time from the
   $4^{th}$ order extension of the Boris integrator (magenta), the Vay integrator (green),
   and the Higuera-Cary integrator (blue) with a step size of $4$ ns (around $0.008$ oscillation period).
\begin{figure}[!htb]
%\begin{figure}[htb]
%   \vspace*{-.5\baselineskip}
   \centering
   \includegraphics*[angle=270,width=200pt]{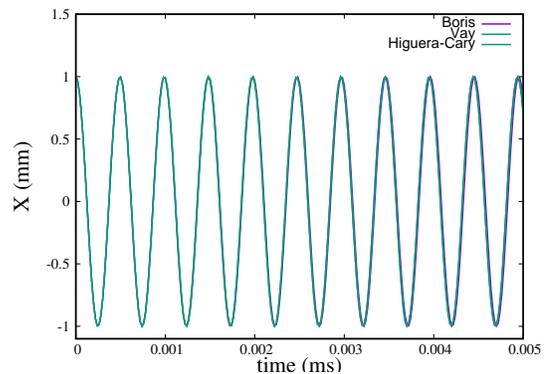}
   \caption{Particle trajectory evolution as a function of time from the
   $4^{th}$ order extension of the Boris integrator (magenta), the Vay integrator (green),
   and the Higuera-Cary integrator (blue) for an electron
   with $50$ MeV kinetic energy.}
   \label{fig3}
%   \vspace*{-\baselineskip}
\end{figure}
It is seen that at beginning all three integrators agree with each other well.
After $2$ ns, the Boris integrator starts to deviate from the other
two numerical integrators due to the cancellation errors from the electric
field and the magnetic field.
Figure~\ref{fig4} shows the relative numerical errors at the end of integration
	   as a function of step size from the
   $4^{th}$ order extension of the Boris integrator (magenta), the Vay integrator (green),
   and the Higuera-Cary integrator (blue) together with a power $4$ polynomial.
It is seen that all three numerical integrators 
converge as $4^{th}$ power of the step size.
However, the $4^{th}$ order extension of the Boris integrator still shows much larger 
relative errors than the other two $4^{th}$ order integrators.
\begin{figure}[!htb]
%\begin{figure}[htb]
%   \vspace*{-.5\baselineskip}
   \centering
   \includegraphics*[angle=270,width=200pt]{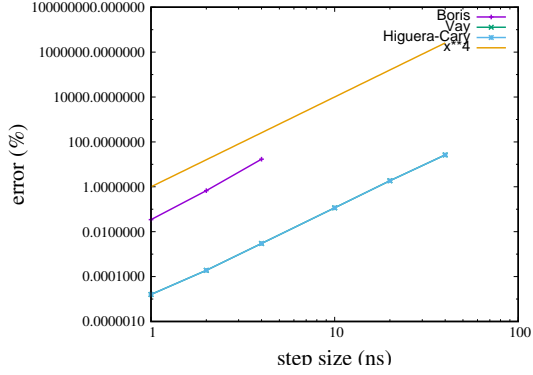}
   \caption{Relative numerical errors at the end of above integration
	   as a function of step size from the
   $4^{th}$ order extension of the Boris integrator (magenta), the Vay integrator (green),
   and the Higuera-Cary integrator (blue) for an electron
   with $50$ MeV kinetic energy. A power $4$ polynomial is also plotted here (orange).}
   \label{fig4}
%   \vspace*{-\baselineskip}
\end{figure}

\begin{figure}[!htb]
   \centering
   \includegraphics*[angle=270,width=200pt]{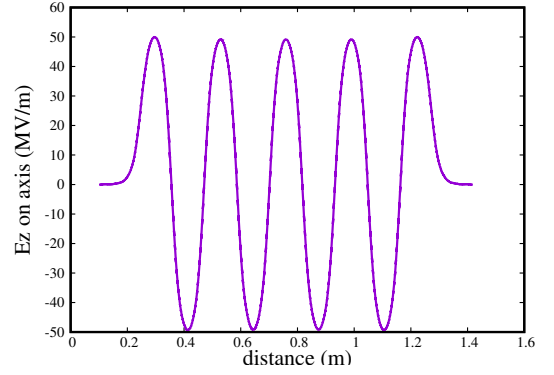}
   \caption{On-axis electric field inside the RF cavity.}
   \label{fld}
\end{figure}

In the second example, we assume that a $10$ MeV electron transports through
a standing wave radio-frequency (RF) cavity with time-dependent electromagnetic fields.
The electromagnetic fields are given as:
\begin{eqnarray}
	E_x  & = & -x \sum_{n=0}^{1}\frac{1}{2(n+1)}e_n'(z)r^{2n}
\cos(\omega t + \theta)  \\
E_y  & = & -y \sum_{n=0}^{1}\frac{1}{2(n+1)}e_n'(z)r^{2n}
\cos(\omega t + \theta)  \\
E_z  & = & \sum_{n=0}^{1}e_n(z)r^{2n}
\cos(\omega t + \theta) \\ 
	B_x  & = & y \frac{1}{\omega} \sum_{n=0}^{1}\frac{1}{2(n+1)}e_n(z)r^{2n}
\sin(\omega t + \theta)  \\
B_y  & = & -x \frac{1}{\omega} \sum_{n=0}^{1}\frac{1}{2(n+1)}e_n(z)r^{2n}
\sin(\omega t + \theta)  \\
B_z  & = & 0
\end{eqnarray}
with $r^2 = x^2+y^2$ and
\begin{eqnarray}
e_{n+1}(z) & = & -\frac{1}{4(n+1)^2}(e_n''(z) + \frac{\omega^2}{c^2}e_n(z))
\end{eqnarray}
where $\omega$ is the RF angular frequency of the cavity, $\theta$ is the initial
driven phase of the cavity, and $e_0(z)$ is the on-axis longitudinal electric field.
In this case, the RF frequency is $1.3$ GHz, and the initial phase is $224$ degree,
and the on-axis electric field $e_0(z)$ is shown in Fig.~\ref{fld}.

\begin{figure}[!htb]
%\begin{figure}[htb]
%   \vspace*{-.5\baselineskip}
   \centering
   \includegraphics*[angle=270,width=200pt]{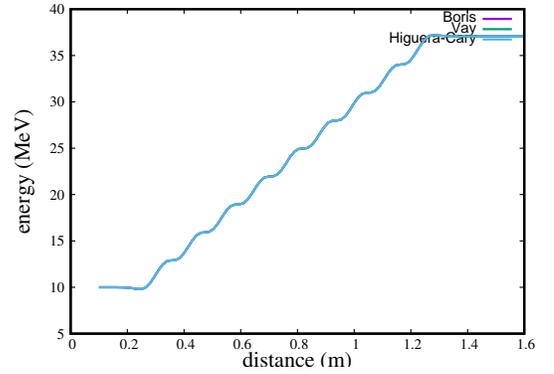}
   \caption{Particle kinetic energy evolution through the RF cavity.}
   \label{fig5}
%   \vspace*{-\baselineskip}
\end{figure}

\begin{figure}[!htb]
%\begin{figure}[htb]
%   \vspace*{-.5\baselineskip}
   \centering
   \includegraphics*[angle=270,width=200pt]{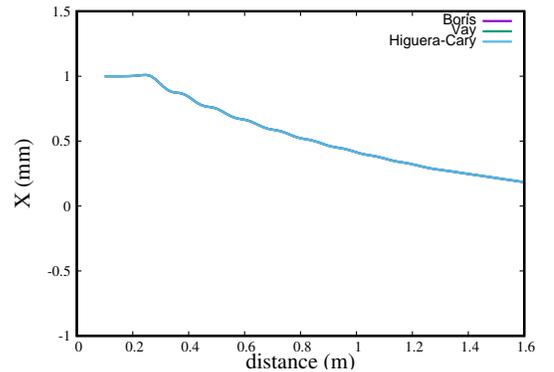}
   \caption{Particle trajector evolution through the RF cavity.}
   \label{fig6}
%   \vspace*{-\baselineskip}
\end{figure}

The electron kinetic energy evolution through the RF cavity is shown in Fig.~\ref{fig5}.
The electron is accelerated from the initial $10$ MeV to the final about $37$ MeV
at the exit of the cavity.
Figure~\ref{fig6} shows the electron trajectory evolution through the cavity. 
The electron is focused from the transverse electromagnetic forces through the cavity.
\begin{figure}[!htb]
%\begin{figure}[htb]
%   \vspace*{-.5\baselineskip}
   \centering
   \includegraphics*[angle=270,width=200pt]{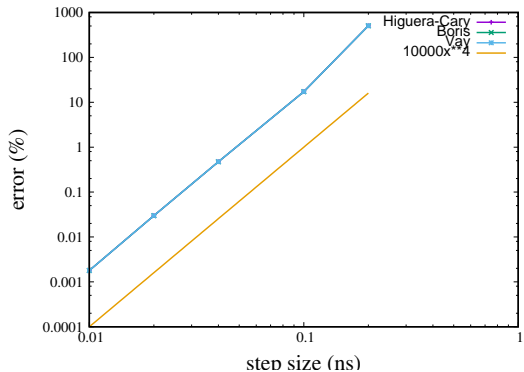}
   \caption{Relative numerical errors at the exit of the cavity
	   as a function of step size from the
   $4^{th}$ order extension of the Boris integrator (magenta), of the Vay integrator (green),
   and the Higuera-Cary integrator (blue).
   A power $4$ polynomial is also plotted here (orange).}
   \label{fig7}
%   \vspace*{-\baselineskip}
\end{figure}
Figure~\ref{fig7} shows the relative numerical errors at the exit of the cavity
	   as a function of step size from the
$4^{th}$ order extension of the Boris integrator (magenta), the Vay integrator (green),
and the Higuera-Cary integrator (blue) together with a plot of the $4^{th}$ 
power polynomial.           
It is seen that in this example, all three $4^{th}$ order numerical
integrators have nearly the same relative errors and converge as $4^{th}$ power
of the step size.

\section{Conclusion and Discussion}

In the paper, three second order, time reversible numerical integrators were 
extended to $4^{th}$ and arbitrary even order accuracy integrators 
following the split-operator
method. These high order numerical integrators have the potential to
significantly
save computational cost with a given numerical error tolerance. The two high
order relativistic integrators can also be used to 
track the charged particle with large relativistic factor in electromagnetic
fields.
These integrators implemented in some modern 
beam dynamics simulation code~\cite{impact-t} will be a useful tool for high brightness electron
beam dynamics study.

The extension to higher order accuracy presented in this paper is not limited
to the above three integrators. The same extension can be applied to 
the other time reversible second order relativistic 
integrators~\cite{qin,petri}, which were noticed by the author after this
work had been done.

\section*{ACKNOWLEDGEMENTS}
Work supported by the U.S. Department of Energy under Contract No. DE-AC02-05CH11231.
This research used computer resources at the National Energy Research
Scientific Computing Center.

\end{document}